\documentclass[aps,prb,twocolumn,superscriptaddress,showpacs]{revtex4}
\usepackage{ae}
\usepackage[T1]{fontenc}
\usepackage[ansinew]{inputenc}
\usepackage{amsmath}
\usepackage{amssymb}
\usepackage{graphicx}
\usepackage{color}
\usepackage[colorlinks]{hyperref}
\usepackage{epstopdf}

\begin{document}

\date{\today}

\title{Dynamics of Current Induced Magnetic Superstructures in  Exchange-Spring Devices }

\author{A. M. Kadigrobov}
\affiliation{Department of Physics, University of Gothenburg,
SE-412 96 Gothenburg, Sweden} \affiliation{Theoretische Physik
III, Ruhr-Universit\"{a}t Bochum, D-44801 Bochum, Germany}
\author{R. I. Shekhter}
\affiliation{Department of Physics, University of Gothenburg,
SE-412 96 Gothenburg, Sweden}
\author{M. Jonson}
\affiliation{Department of Physics, University of Gothenburg,
SE-412 96 Gothenburg, Sweden} \affiliation{SUPA, Institute of
Photonics and Quantum Sciences, Heriot-Watt University, Edinburgh EH14 4AS, Scotland, UK}
\affiliation{Department of Physics, Division of Quantum Phases and Devices, Konkuk University, Seoul 143-701, Republic of Korea}


\begin{abstract}
Thermoelectric manipulation of the magnetization of  a magnetic layered stack in which  a low-Curie temperature magnet is sandwiched between two strong magnets (exchange spring device) is considered. Controllable Joule heating produced by a current flowing in the plane of the magnetic stack  (CIP configuration) induces a spatial magnetic and thermal structure along the current flow --- a magneto-thermal-electric domain  (soliton). We show that such a structure can experience oscillatory in time dynamics if  the magnetic stack is incorporated into an electric circuit in series with an inductor. The excitation of these  magneto-thermionic oscillations follow the scenario  either of "soft" of  "hard"  instability: in the latter case  oscillations arise if the  initial perturbation is large enough. The frequency of the temporal oscillations is of the order of $10^5 \div 10^7 s^{-1}$ for current densities $j\sim 10^6 \div 10^7 A/cm^3$.
\end{abstract}

\maketitle

\section{Introduction}
The electric control of magnetization on the nanometer length scale became a subject of intensive study after the seminal suggestion by Slonzewsky\cite{Slonczewski,Berger} to use the spin torque transfer (STT) technique, where a current of spin-polarized electrons is injected into a magnetic material. Even though the magnetic tuning induced by the exchange interaction between the injected electrons and those that make the material magnetically ordered can only be achieved in a very small region, large current densities are needed to get a significant STT effect. Such current densities can be reached  in electric point contacts where values  of the order of $10^8 \div 10^9$~A/cm$^3$ have been achieved and the STT effect was detected.\cite{Rippard,Yanson} By a further increase of the current thermal Joule heating of the material becomes unavoidable. However, the fact that such heating can be precisely controlled by the bias voltage allows thermoelectrical manipulation of the magnetization direction based on the orientational phase transition in layered  magnetic structures predicted in Ref.~\onlinecite{Jouleheating,JAP,s-shaped}. It was shown there that in such structures with different Curie temperatures of the layers
 there can exist  a finite temperature interval inside which the angle of the relative orientation of the layer magnetization $\Theta$
 depends  on temperature, $\Theta=\Theta(T)$, changing reversibly  with a change of temperature in the whole interval of angles from the parallel to the antiparallel orientation. As a result, by controlling the Joule heating by the bias voltage $V$ one may smoothly and reversibly control the  relative  magnetization angle  in the whole angle interval $0\leq \Theta[T(V)]\leq \pi$.
 There was also  suggested a thermal-electronic oscillator device  based on  this magneto-thermal effect in  magnetic exchange spring  structures.

 In paper~\onlinecite{MTED}, for  exchange-spring magnetic stacks\cite{Davies} with an in-plane electric current flow (CIP) (see Fig.~\ref{noflip}),  it was predicted and investigated  a possibility of generation of  magnetic spatial superstructures (domains and solitons)  the appearance of which and their  parameters can be  controlled by the bias voltage. In the  present paper  we study  temporal dynamics  of such spatial structures in the case that the exchange-spring magnetic stack is incorporated in an external circuit in series with an inductor \cite{Chiang1}. Such oscillations can be understood as  a realization of the bi-stability of an N-shaped IV curve of the device in the regime at which a large enough inductance of the circuit determines slow temporal variations of the current.

The structure of the paper is as follows. In Section \ref{n-shapedIVC} we present in short some peculiar  features of the dependence of the magnetization orientation angle in the stack on temperature $T$ and some properties of the magnetic stack with a magnetic-thermal-electric domain (MTED) inside it  needed for  further considerations. In Section \ref{MED}  we show that MTED loses its stability if the stack is incorporated  in an external circuit with a large enough inductance. We develop there an adiabatic theory which allows an analytical  description of the  MTED dynamics.

\section{The magnetic-thermal-electric domain in the stack under Joule heating \label{n-shapedIVC}}

\begin{figure}
 \centerline{\includegraphics[width=0.7\columnwidth]{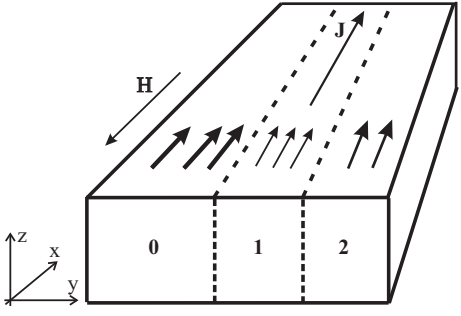}}
  \caption{ Orientation of the magnetic moments in
  a  stack of three ferromagnetic layers
(0, 1, 2); the magnetic moments in layers 0, 1, and 2 (shown with short arrows) are coupled
by the exchange interaction thus building an exchange spring
tri-layer; $H$ (shown with the arrow outside the stack)  is an external magnetic field directed antiparallel
to the magnetization in layer 0. The current $J$ (shown with the long arrow) flows in the layer
plane (that is along the x-axis).}
 \label{noflip}
 \end{figure}

 The system under consideration has  three ferromagnetic layers   in which two strongly
 ferromagnetic layers 0 and 2 are exchange coupled through a weakly ferromagnetic spacer
 (layer 1) as is shown in Fig.~\ref{noflip}. We assume that the Curie temperature $T_c^{(1)}$
 of layer 1
is lower than the Curie temperatures $T_c^{(0,2)}$ of layers 0, 2;
we also assume the magnetization direction of layer 0 to be fixed;
this stack is under an external magnetic field $H$ directed
opposite to the magnetization of layer 0. We require this
magneto-static field  to be weak enough so that at low
temperatures $T$ the magnetization of layer 2 is kept parallel to
the magnetization of layer 0 due to the exchange interaction
between them via layer 1. At $H=0$ and $T>T_c^{(1)}$ this tri-layer is similar to the spin-flip "free layer" widely used in memory device application \cite{Worledge}.
The stack is incorporated into an external circuit along which the
total current $J$ flows through the cross-section of the layers:
\begin{eqnarray}\label{current}
J= \Big[\frac{1}{R\big(\Theta\big)} +\frac{1}{R_0} \Big]V
\end{eqnarray}
where $R\big(\theta\big)$ and $R_0$ are the magnetoresistance  and
the angle-independent resistance of the stack, $\Theta$ is the
angle between the magnetization directions of layers 0 and 2, $V$
is the voltage drop across the stack.

In paper \cite{JAP}, it was shown that parallel orientations of
the magnetization in layers 0, 1 and 2 becomes unstable if the
temperature exceeds some critical temperature $T_c^{\rm(or)} <
T_c^{(1)}$. The magnetization direction in layer 2 smoothly tilts
with an increase of the stack temperature $T$ in the temperature
interval $T_c^{\rm(or)} \leq T \leq T_c^{(1)}$. The dependence of
the  equilibrium tilt angle $\Theta$ between the magnetization
directions of layers 0 and 2 on $T$ and the magnetic filed $H$ is
determined by the equation \cite{JAP}
\begin{eqnarray}\label{theta2}
&\Theta&=D(H,T)\sin{\Theta}, \hspace{0.2cm} T < T_c^{(1)} \nonumber \\
&\Theta&=\pm \pi, \hspace{1.9cm} T \geq T_c^{(1)}
\end{eqnarray}
where
\begin{equation}
D(H,T)=\frac{L_1 L_2H M_2(T)}{4 \alpha_1 M_1^2(T)}\approx D_0(H)
\frac{T_c^{(1)}}{T_c^{(1)}-T}.
 \label{D}
\end{equation}
and
\begin{equation}
D_0(H)=\frac{\mu_B H}{k_B T_c^{(1)}}
\Bigl(\frac{L_1}{a}\Bigr)\Bigl(\frac{L_2}{a}\Bigr)
 \label{D_0}
\end{equation}
Here   $L_1, \; M_1(T) $ and $L_2, \; M_2(T)$ are the widths  and
the magnetic moments of layers 1 and 2, respectively; $\alpha_1
\sim I_1/ a M_1^2(0)$ is the exchange constant, $I_1$ is the
exchange energy in layer 1, $\mu_B$ is the Bohr magneton, $k_B$ is
Boltzmann's constant,  and $a$ is the lattice spacing.     $D(H,T)$ is a dimensionless
parameter that determines the efficiency of the external magnetic in the misorientation effect under the consideration. It is a ratio between the energy of magnetic layer 2 in the external magnetic field and the energy of the indirect exchange between layers 0 and 2 (see Fig.~ \ref{noflip}).

The critical temperature of the orientation transition \cite{Giovanni} $T_c^{(or)}$ is determined by the condition $D(T)=1$ and is equal to
\begin{eqnarray}
T_c^{\rm(or)}= T_c^{(1)}\left(1- \frac{\delta T}{T_c^{(1)}}
\right), \hspace{0.2cm}
 \frac{\delta T}{T_c^{(1)}}=D_0(H); \nonumber\\
 \label{Torient}
\end{eqnarray}
Taking experimental values (see Ref.~\onlinecite{JAP}) $L_1=30$ nm, $L_2=12$ nm, $T_c^{(1)}=373$K,   and magnetic field $H=10 \div 47 $~Oe one finds $D_0 = 0.1 \div 0.36$  and the temperature interval $\delta T = T_c^{(1)} - T_c^{(or)} =D_0 T_c^{(1)} \approx 37.3 \div 134$~K.

If the stack  is Joule heated  by  current $J$ its temperature
$T(V)$   is determined by  the heat-balance condition
\begin{equation}
JV=Q(T), \hspace{0.2cm}J  =V/R_{\rm eff}(\Theta),
 \label{heat}
\end{equation}
where
\begin{equation}
R_{\rm eff}(\Theta)=\frac{R(\Theta)R_0}{R(\Theta)+R_0},
 \label{Reff}
\end{equation}
 and Eq.~(\ref{theta2}) which determines  the temperature dependence
of $\Theta[T]$. Here  $V$ is the voltage drop across the stack,
$Q(T)$ is the heat flux flowing from the stack and $R_{\rm eff}(\Theta)$ is
the total stack magnetoresistance. Here and below we neglect the
explicit dependence of the magnetoresistance on $T$ considering the main mechanism of the stack resistance to be  the elastic scattering of electrons by impurities.
Below we drop the subscript "eff" at the magnetoresistance symbol $R_{\rm eff}$.
On the other hand, we consider the temperature
changes caused by the Joule heating only in a narrow vicinity of
$T_c^{(1)}$ which is sufficiently lower than both the critical
temperatures $T_c^{(0,2)}$ and the Debye temperature.

\begin{figure}
 \centerline{\includegraphics[width=0.85\columnwidth]{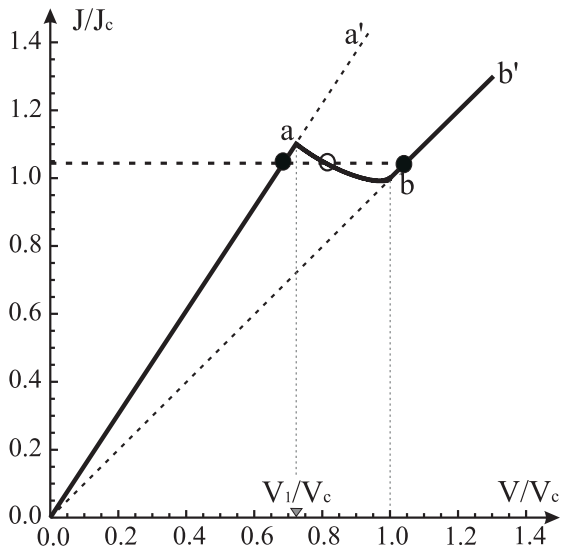}}
\vspace{0 mm} \caption{ Current-voltage characteristics  of
the magnetic stack of Fig.~\ref{noflip} in which the  magnetization directions in the layers are homogeneously distributed along the $x-$direction (that is along the stack).
It was calculated for
$R(\Theta)=R_{+}-R_{-} \cos{\Theta}$, $R_-/R_+ =0.2$, $D_0=0.2$;
$J_c=V_c/R(\pi)$. The branches  $0-a$ and $b-b'$ of the IVC
correspond to parallel and antiparallel orientations of the stack
magnetization, respectively (the parts $a-a'$ and $0-b$ are
unstable); the branch $a-b$ corresponds to  $0\leq
\Theta[T(V)]\leq \pi$.
}
 \label{IVC}
\end{figure}

Equations~(\ref{heat}) and (\ref{theta2}) define the IVC of the
stack
\begin{equation}
J_0(V)=\frac{V}{R[\Theta(V)]}, \label{IVC0}
\end{equation}
where $\Theta(V)\equiv \Theta[T(V)]$.
The differential conductance
$G_{diff}\equiv dJ/dV$ is negative \cite{JAP} if
\begin{equation}
\frac{d}{d \Theta}\frac{(1-{\bar D}\sin{\Theta}/\Theta)}{R(\Theta)
}<0
 \label{diffGinequality}
\end{equation}
where ${\bar D}=(T/Q)(dQ/dT)D_0 \approx D_0$
In this case the current-voltage characteristics of the stack
(IVC) is N-shaped as is shown in Fig.~\ref{IVC}.

In  paper Ref.~\onlinecite{MTED} it was shown that    a homogeneous in space distribution of the
magnetization direction, temperature and electric field along the
spring-type magnetic stack  becomes unstable and a
magneto-thermal-electric domain (MTED) spontaneous arise \cite{domain} if the electric current flows in the plane of the layers (CIP-cofiguration) and the length  of the stack exceeds the  critical length
\begin{eqnarray}\label{Lcritical}
L_c \approx  \sqrt{\frac{\kappa }{r}\frac{T_c^{(1)}}{ \rho (0)
j^2}}
\end{eqnarray}
where $\kappa$ is the thermal conductivity, $\rho(\Theta)$ is the magnetoresistivity and $r=(\rho(0)-\rho(\pi))/(\rho(0)+\rho(\pi))$. Parameters of the MTED which arises in the stack, depend on the stack length and the voltage drop across it that allows to create and control magnetic structures in magnetic devices.  Using Eq.~(\ref{Lcritical}) and the
Lorentz ration $\kappa/\sigma =\pi^2k_B^2T/3 e$ ($\sigma=\rho^{-1}$
and $e$ is the electron charge) one finds that
\begin{eqnarray}\label{Lcestimations}
L_c \sim \frac{\pi }{\sqrt{r}}\frac{k_B T_c^{(1)}}{e \rho(0)
j_c} \sim 10  \mu m
\end{eqnarray}
for a realistic experimental situation \cite{torquecurrent} $r\sim 0.1\div 0.3$,
$T_c^{(1)}\sim 10^2$K, $\rho (0)\sim 10^{-5}\Omega$cm, $j \sim
10^6\div 10^7 A/cm^2$.

The magnetic-thermal-electric domain (MTED) which determines the space distributions of the temperature $T_d(x)$ and the magnetization direction angle $\Theta(x)$ which spontaneously arise in the magnetic stack, satisfies the following equation  \cite{MTED}:
\begin{eqnarray}
\kappa \frac{d^2T_d}{dx^2} =j^2\rho [ \theta(T_d)]-Q(T_d)/\Omega_{st}
\label{MEDEQ}
\end{eqnarray}
in which current $j$ is found from the equation
\begin{eqnarray}
j\left<\rho[\Theta(T_d)] \right>=\frac{V}{L}
\label{current-bias}
\end{eqnarray}
supplemented with the periodic boundary condition $T_d(x+L)=T_d(x)$ ($x$ is the coordinate along the stack, $L$ and $\Omega_{st}$ are the length and  the volume of the stack, respectively; the brackets $<...>$ indicate an average
over $x$ along the whole stack of the length  $ L$).

\begin{figure}
 \centerline{\includegraphics[width=0.85\columnwidth]{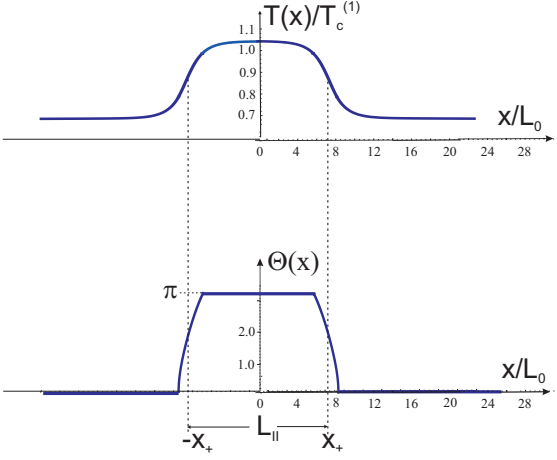}}
\vspace{0 mm} \caption{The coordinate dependence of the temperature, $T(x)$, and the magnetization direction, $\Theta(x)$, in the magnetic stack with a magneto-thermal-electric domain inside it, $L_{II}$ is the length of the "hot" part of the MTED which is defined in such a way that the length of the "cold" part is $L_{I}=L-L_{II}$  Calculated are made for
$R(\Theta)=R_{+}-R_{-} \cos{\Theta}$, $R_-/R_+ =0.2$, $D_0=0.2$;
$J/J_c=1.0265$, $J_c=V_c/R(\pi)$, $L_0 =\sqrt{j_c^2\rho(\pi)/\kappa T_c^{(1)}}\sim L_c$.
}
 \label{T(x)}
\end{figure}

The space distribution of the temperature $T_d(x)$ and the magnetization direction angle $\Theta(x)$  in a long stack ($L\gg L_c$)  with a MTED inside it  is shown in Fig.~\ref{T(x)}.

In this limit ($L\gg L_c$) the MTED is of a trapezoid form with two planar segments of  the length $L_I$  and $L_{II}$ ($L_{I}+L_{II} =L $) with two transition regions of the width $\sim L_c$ (see Fig.~\ref{T(x)}).

Neglecting
the contribution of the transition region in the total voltage drop across the stack,   the dependence of  the lengths  $L_I$ and $L_{II}$  on the  voltage drop $V$ across the stack can be written as
\begin{eqnarray}\label{L2}
L_I \approx  \frac{V/L- {\cal E}_{I}}{2r\rho_+j},\; L_{II} \approx  \frac{ {\cal E}_{II}-V/L}{2r\rho_+j}
\end{eqnarray}
where ${\cal E}_{I}=\rho(0)j$, ${\cal E}_{II}=\rho(\pi)j$ and $r=\rho_{-}/\rho_{+}$, $\rho_{\pm}=(\rho(\pi)\pm \rho(0))/2$. In this approximation these formulas are valid in the range of $V$ where $L_{I,II}\geq 0$.
  The current $J= S j$ and the voltage $V={\cal E} L$ are coupled via  current-voltage characteristics $j=j_d(V/L)$ for the stack with MTED inside it. This dynamic IVC  has a form close to a plateau in which the current $j$ coincides with the stabilization current \cite{MTED} $j_0$ to an accuracy exponential in $L/L_c$.
The dynamic CVC can be  presented in an  implicit form as
\begin{eqnarray}\label{dynamicCVC}
j-j_0 = J_I \exp\Big\{- \frac{{\cal E}-\rho(0)j}{r(j_0\rho_{+})}\frac{L}{L_{0}}\Big\}\nonumber  \\
-J_{II} \exp\Big\{\frac{{\cal E}-\rho(\pi)j}{r(j_0\rho_{+})}\frac{L}{L_{0}}\Big\}
\end{eqnarray}
Here   ${\cal E}=V/L$,  $L_0=\sqrt{f^{'}_T(T_c^{(1)})/\kappa}\sim L_c$ and the constants $J_{I,II}$ are of the order of $ j_{0}$ while $f{'}_T=\partial f/\partial T$.

\section{Time Evolution of Magneto-Electric-Thermal Domain  in Magnetic Stack \label{MED}}

Here we assume that the  magnetic stack is placed in series with an inductor. We also assume the bias-voltage regime that is
 the resistance
 of the external circuit in which the magnetic stack is incorporated can be neglected in
 comparison with that of the stack. In this case, taking into account that the temperature $T(x,t)$  (here  $t$ is
the time) satisfies the equation of continuity of heat flow (see, e.g., Ref.~\onlinecite{Landau}),   one obtains the following set of basic  equations of the problem:
 \begin{eqnarray}
&&c_v\frac{\partial T}{\partial t}
- \frac{\partial}{\partial x}\Big(\kappa(T) \frac{\partial T}{\partial x}\Big)=-f(T,j)\nonumber\\
 &&\frac{{\cal L} S}{L}\frac{d j(t)}{d t}+\left<\rho \left[\Theta (T)\right]\right>j(t)=\frac{V}{L}
\label{heateq}
\end{eqnarray}
where
\begin{eqnarray}
f(T,j)=Q(T)/\Omega_{st}-j^2(t)\rho [ \theta(T)].
\label{heateq1}
\end{eqnarray}
Here the dependence $\Theta(T)$ is given by Eq.~(\ref{theta2}). In
the above equation   $j(t)=J(t)/S$ is the current density which is independent of
$x$ due to the condition of local electrical neutrality, $S$ is the cross-section area of the stack and $L$ is its length, $c_v$ is
the heat capacity per unit volume,  $Q(T)$ is the
heat flux flowing from the stack, $\Omega_{st}$ is its volume and
$\rho[\theta]=R[\theta]S/L$ is the stack magneto-resistivity
(see Eq.~(\ref{Reff})),  ${\cal L}$ is the inductance,  and  $V$ is the bias voltage; the definition of the brackets <...> is the same as in Eq.\ref{current-bias}.

The boundary condition to Eq.~(\ref{heateq}) is the continuity of
the heat flux at the both ends of the stack (which is coupled to
an external circuit with a fixed voltage drop $V$ over the stack).
We shall not write down the expression for this condition, since   the domain dynamics  does not sufficiently depend on it.
Instead, for the sake of simplicity we use the periodicity
condition $T(x+L,t)=T(x,t)$.

The set of equations Eq.~(\ref{heateq}) always has a steady-state solution which represent a magnetic-thermal-electric domain $T_d(x)$ which satisfies Eq.~(\ref{MEDEQ}).
A study of the  stability of this solution carried out using a linearized system Eq.~(\ref{heateq}) shows that the MTED loses its stability if the inductance ${\cal L}$  exceeds some critical value ${\cal L}_{c1}\sim \tau_TR_+  (L/L_c) \exp{L/L_c} $ where $\tau_T \sim C_V T/RJ^2$ is the characteristic evolution  time of the temperature\cite{footnote2}. Therefore, for the case $L\gg L_c$ this instability develops under conditions that   the characteristic time of the current evolution $\tau_{{\cal L}}\sim {\cal L}/R$ is much longer than the temperature evolution time $\tau_T$ (their ratio is $\tau_{{\cal L}}/\tau_T \sim (L/L_c)\exp{L/L_c}$. Then  any small fluctuation develops in such a way that after a time $t\gg \tau_T$ the domain recovers its initial trapezoidal form and only its length $L_{II}$ varies slowly with time $t$. This fact makes it possible to obtain a reduced description of the nonlinear dynamics of a MTED as follows. We seek the solution of the first equation in Eq.~(\ref{heateq})  in the form

\begin{eqnarray}\label{AdiabaticDynamic}
T(x,t)=T_d(x,j_d(t))+T_1(x,t)
\end{eqnarray}
where $T_d(x,j_d)$ is the solution of the equation which is obtained from  Eq.~(\ref{MEDEQ}) by the substitution $j\rightarrow j_d(t)$; the parameter $j_d$ governing the length of the domain $L_{II}(j_d)$ is a slow function of the time $t$ which has to be found; $T_1(x,t)$ is a  correction  which is small with respect to $T_d$.
Inserting Eq.~(\ref{AdiabaticDynamic}) into the first equation in Eq.~(\ref{heateq}) and linearizing it one gets the following equation for $T_1$:
\begin{eqnarray}\label{Hamiltonian1}
\hat{H}T_1= -c_V\frac{\partial T_d}{\partial j_d}\frac{dj_d}{d t} +\Big(j^2-j^2_d\Big)\rho(T_d)
\end{eqnarray}
where   $\hat{H}$ is a Hermitian operator with a periodic boundary condition:
\begin{eqnarray}\label{Hamiltonian2}
\hat{H}=-\kappa \frac{d^2}{d^2 x}+
\Big(-j_d^2
\rho_T^{'}(T_d)
 +Q_T^{'}(T_d)/\Omega_{st}\Big)
\end{eqnarray}
(here $\rho(T) \equiv \rho[\Theta(T)]$ and $F^{'}_T = d F/d T$).
The requirement of the smallness of the ratio $|T_1/T_d|\ll 1$ allows to find the needed equations describing the adiabatic MTED evolution.

The  solution of  Eq.~(\ref{Hamiltonian1}) is found in the form of expansion
\begin{eqnarray}\label{Texpansion}
T_1(x,t)= \sum_{\nu} A_{\nu}(t)\Psi_\nu(x)
\end{eqnarray}
where $\Psi_\nu$ are the eigenfunctions of $\hat{H}$ which satisfy the Sturm-Liouville equation
\begin{eqnarray}\label{eigenfunctionequation}
\Big[-\kappa \frac{d^2}{d^2 x}
-j_d^2
\rho_T^{'}(T_d)
 +Q_T^{'}(T_d)/\Omega_{st}\Big]\Psi_\nu=\lambda_\nu \Psi_\nu
\end{eqnarray}

Multiplying the both sides of Eq.~(\ref{Hamiltonian1}) by $\Psi_\nu$ from the left and averaging over the period $L$  one finds
\begin{eqnarray}\label{Aequation}
\lambda_\nu A_\nu  =\Big<\Big( -c_V \frac{\partial T_d}{\partial j_d}\frac{d j_d}{dt} +(j^2-j_d^2)\rho(T_d)\Big)\Psi_\nu \Big>
\end{eqnarray}
where $\partial T_d/\partial j_d$ in terms of eigenfunctions $\Psi_\nu$ is presented in Appendix \ref{eigenvalues}, Eq.~(\ref{dT/dj}).

As the eigenvalue $\lambda_0$ is exponentially small (see Eq.~(\ref{lambda0})), from Eq.~(\ref{Hamiltonian1}) it follows that the requirement $|T_1/T_d|\ll 1$ is satisfied only when  the right-hand side of the equation  with $\nu =0$ is equal to zero:
\begin{eqnarray}\label{reducedDynamic1}
\Big<\Big( -c_V \frac{\partial T_d}{\partial j_d}\frac{d j_d}{dt} +(j^2-j_d^2)\rho(T_d)\Big)\Psi_0 \Big>=0.
\end{eqnarray}
Using Eq.~(\ref{PsiAntisymmetric}) one sees that the right-hand side of the equation with $\nu=1$  is equal to zero and hence the factor $\lambda_1=0$ in the left-hand side of it does not violate the above-mentioned requirement.

Taking into account the second equation in Eq.~(\ref{heateq}) and   Eq.~(\ref{reducedDynamic1}) together with Eqs.(\ref{TderivativeJeq}, \ref{lambda01}) one finds a set of equations which describes the temporal dynamics of a MTED:
\begin{eqnarray}\label{reducedDynamics}
\frac{d j_d}{d t}&=&\omega_0(j_d)\frac{j^2-j_d^2}{2j_0}, \nonumber\\
\bar{{\cal L}}\frac{d j}{d t}&=&V/L -\langle\rho(T_d(x,j_d))j
\end{eqnarray}
Here $\bar{{\cal L}}=S {\cal\ L}/L$ and $\omega_0=-\lambda_0/c_V$ (here $\lambda_0$ is defined by Eq.~(\ref{lambda01})) may be written as
\begin{eqnarray}\label{omega0}
\omega_0^{-1} (j)=-\tau_0\frac{2 r}{\rho_+}\frac{d{\cal E}_d}{dj}
\end{eqnarray}
where the constant $\tau_0$ is
\begin{eqnarray}\label{tau}
\tau_0=\frac{c_VL^2\langle(dT_d/dx)^2\rangle}{8 j^2\int_{T_{min}}^{T_{mâx}}\rho(T)dT}\sim\tau_T,
\end{eqnarray}
being of the order of the characteristic evolution  time  of the temperature $\tau_T$, and ${\cal E}_d$ is defined in Eq.(\ref{current-bias1}).

Throughout the  range of the existence of a trapezoidal MTED one has $|j-j_0|\ll j_0$ and the set of equations Eq.~(\ref{reducedDynamics}) can be reduced to an equation for the electric field which is coupled to the current $j_d$  via the voltage-current characteristic    ${\cal E}={\cal E}_d(j_d)$ (see Eqs.(\ref{current-bias},\ref{current-bias1})). Differentiating the first equation in Eq.~(\ref{reducedDynamics}) multiplied by $\omega_0$ with respect to $t$,  and inserting in the resulting expression the above-mentioned electric field ${\cal E}$  as a new variable together with the second equation in Eq.~(\ref{reducedDynamics})
one obtains the following equation in the form of a linear oscillator with a nonlinear nonconservative term:
\begin{eqnarray}\label{OscillatorEquation}
\ddot{{\cal E}}+
\Big(\frac{{\cal E}}{ \bar{{\cal L}}j_0}
+\frac{\rho_+}{\tau_0} \frac{d j_d}{d {\cal E}}\big)\dot{{\cal E}}+\frac{\rho_+}{ \bar{{\cal L}}\tau_0}\left({\cal E}-V/L\right)=0.
\end{eqnarray}
Here $j_d(\cal E)$ is the CVC of the magnetic stack with a MTED.

The static points of Eq.~(\ref{OscillatorEquation}) and the second equation in Eq.~(\ref{reducedDynamics}) correspond to the domain solution  ${\cal E}=V/L$ and $j=j_d(V/L)$. As $d j_d/d {\cal E} <0$, in the range of inductance
$ \bar{{\cal L}}>  \bar{{\cal L}}_{c1}$ the factor in front of $\dot{{\cal E}}$ is negative and a domain is absolutely unstable.  Here the critical inductance is
\begin{eqnarray}\label{Lcritical1}
\bar{{\cal L}}_{c1}\approx -\tau_0
\Big(\frac{d j_d}{d { \cal E}}\Big)^{-1}_{{\cal E}={\cal E}_{inf}}\sim \tau_0 \rho_+ \exp(L/L_c)
\end{eqnarray}
where ${ \cal E}_{inf}$ is the electric field at which $- dj_d/d {\cal E}$ has a maximum that is at which $dj^2_d/d {\cal E}^2=0$. In this region the voltage drop across the magnetic stack $V(t)={ \cal E}(t)  L$  oscillates and hence  the length of the domain $L_{II}(t)$ (see Eq.~(\ref{L2})) oscillates with an amplitude that increases in time. At the time when $L_{II}$ reaches either values $0,L$ (depending on the initial fluctuation) the domain disappears and the sample becomes homogeneous with temperatures $T_{min}$ or  $T_{max}$, respectively. Further temporal  evolution of the system is described by the set of equations Eq.~(\ref{heateq}) an analysis of which for the case $\tau_T/\tau_{{\cal L}}\ll 1$ shows that the system  exhibits stable large-amplitude spontaneous oscillations of the temperature $T$, current current $J$, magnetization direction $\Theta(T)$,  and the voltage drop across the stack $\tilde{V}=J R[\Theta(T)]$, the oscillations being the same as those predicted in Ref.~\onlinecite{JAP} for a magnetic stack in the absence of MTED.
\begin{figure}
\centerline{\includegraphics[width=0.85\columnwidth]{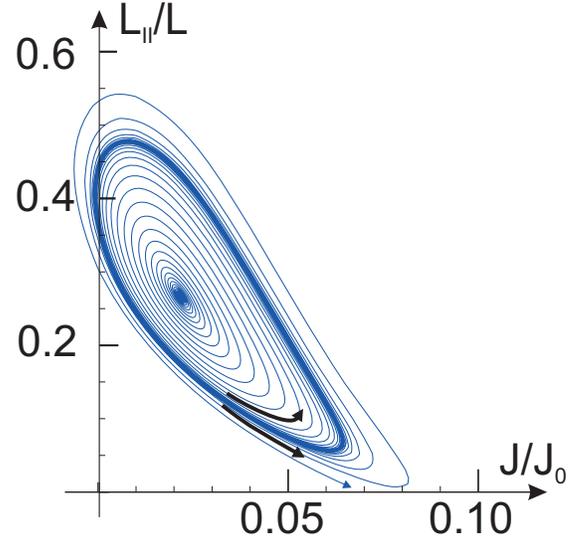}}
\vspace{0 mm} \caption{ Temporal evolution of the length of MTED $L_{II}(t) $  for the case ${\cal L}_{c1}> {\cal L}>{\cal L}_{c2}$. The unstable limiting cycle (shown with a thick line) separate the initial the phase plane into two regions: any initial state inside the limiting cycle develops into the length of the steady stationary MTED (shown with a dot) while an initial   outside it results in oscillations of the MTED length with an increasing in time amplitude until the MTED disappears, that is  $L_{II}$ reaches either $L_{II}\approx 0$ or  $L_{II}\approx L$. Calculations are made   for
  $R_-/R_+=0.2$,  $D_0=0.2$ and $\tau_T/\tau_{{\cal L}}=0.1$ where $\tau_T$ and $\tau_{{\cal L}}$ are the characteristic times of the temperature and current developments, respectively; $J_0$ is the stabilization current.
}
 \label{limlength}
\end{figure}

\begin{figure}
  \centerline{\includegraphics[width=0.85\columnwidth]{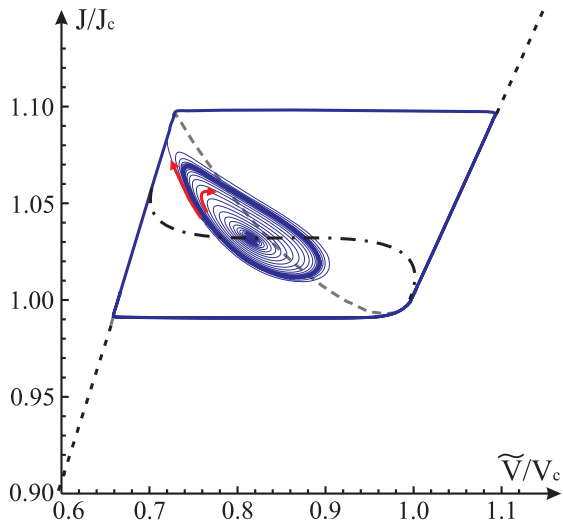}}
  \caption{Temporal evolution  of the current $J(t)$ and voltage drop ${\tilde V}(t)$  in a magnetic stack with a magnetic-thermal-electric domain (MTED) inside it for the case ${\cal L}_{c1}> {\cal L}>{\cal L}_{c2}$. The steady CVC of a homogeneous stack  CVC with a MTED are shown with a dashed  and dashed-dotted lines, respectively.
  The stack is in a bistable state: depending on the initial conditions the system goes either to            MTED        a stable steady MTED (which  is shown with a dot) or goes to a stable limiting cycle (the largest closed curve) corresponding to spontaneous  oscillations  of the current $J(t)$, voltage drop $\tilde{V}(t)$,
  temperature $T(t)$, and the magnetization direction $\Theta(t)$, the stack being in a homogeneous  state (in which the MTED has disappeared). There is an unstable limiting cycle that  separates the initial states which develop  either to the MTED or the the oscillations as is shown by two black arrows.
  Calculations are made   for
  $R_-/R_+=0.2$,  $D_0=0.2$ and $\tau_T/\tau_{{\cal L}}=0.1$ where $\tau_T$ and $\tau_{{\cal L}}$ are the characteristic times of the temperature and current developments, respectively;
$J_c=V_c/R(\pi)$.}
   \label{limIV}
  \end{figure}

Investigation of  stability of the domain solution of Eq.~(\ref{OscillatorEquation}) with the Poincare method (see, e.g., Ref.~\onlinecite{Andronov}) shows that at $\delta {\cal L}={\cal L}_{c1}-{\cal L}\ll {\cal L}$ there is an unstable limiting cycle in the phase plane $({\cal E}, \dot{{\cal E}})$  the radius of which $K\propto \sqrt{\delta {\cal L}/ {\cal L}_{c1}}$ and hence it increases with a decrease of the inductance. From here and from the fact that at ${\cal L}=0$ the domain is absolutely stable  follows an existence of a second critical value of the inductance ${\cal L}_{c2}$ which limits the interval of the hard excitation of oscillations. Therefore, the range of values of the inductance ${\cal L}_{c1} > {\cal L}>{\cal L}_{c2}$ is the region in which a stable MTED, an unstable limiting cycle, and a stable limiting cycle coexist.
For this case, temporal developments  of the MTED length $L_{II}$, current $J$ and voltage drop $\tilde{V}$ across the magnetic stack are shown in Fig.~\ref{limlength} and Fig.~\ref{limIV}.

\section{Conclusion \label{conclusion}}
We have considered the temporal evolution of magneto-thermo-electric domains which  spontaneously arise in an exchange-spring magnetic stack with the electrical current flowing along the layers (CIP configuration). For the case that such a stack is incorporated in an electric circuit in series with an inductor we have found critical values of the inductance at which the magneto-thermal domain loses its stability. We  have developed a perturbation theory in an adiabatic parameter (which is the ratio between the temperature and current characteristic evolution times) theory which allows to investigate the evolution of this instability into an oscillatory in time regime. The excitation of such magneto-thermionic oscillations follow the scenario of either ``soft"  of a ``hard"  instability; in the former case any perturbation results in a spontaneous transition to the oscillatory regime while in the latter case  oscillations only appear if the initial perturbation is large enough. The frequency of the temporal oscillations of the magnetization direction in the stack is of the order of $10^5 \div 10^7$~Hz.

\section{Acknowledgement.}
 Financial support from the EC (FP7-ICT-FET Proj.
No. 225955 STELE), the Swedish VR, and the Korean WCU program
funded by MEST/NFR (R31-2008-000-10057-0) is gratefully
acknowledged.

\appendix

\section{}\label{eigenvalues}

According to Ref.~\onlinecite{MTED} $T_d(x)$ can be written as
\begin{eqnarray}\label{trapezoid}
T_d(x)=\vartheta(x+x_+)+\vartheta(x_+-x)-T_{max}
\end{eqnarray}
where the function $\vartheta(x)$ is a domain-wall type solution of Eq.~(\ref{MEDEQ}) at $j=j_0$ and $L\rightarrow \infty$ which has the following asymptotic
\begin{eqnarray}\label{AsymptoticBehavior}
\lim_{x\rightarrow -\infty}\vartheta(x) =  T_{min} \nonumber \\
\lim_{x\rightarrow \infty} \vartheta(x)= T_{max}
\end{eqnarray}
Here    $\pm x_{+}$ are the points of deflections of the curve $d T_d(x)/x$ so that $L_{II} = x_+ -(-x_+)=2x_+$ is approximately the length of the "hot" section of a trapezoidal  MTED  having the maximal temperature $T_{max}$  and $L_I =L-L_{II}$ is the length of its  "cold" section having  the minimal temperature $T_{min}$ (see Fig.~\ref{T(x)}).

As one sees from Eq.~(\ref{eigenfunctionequation}) the function  $$U(x)= -j_d^2
\rho_T^{'}(T_d)
 +Q_T^{'}(T_d)/\Omega_{st}$$ in the former equation consists of two symmetrical "potential" wells of the width $\sim L_c$, separated by a barrier of the width of  order of $L_{II}$. In this case (see, e.g., Ref.~\onlinecite{LandauQM}) tunneling between  the wells results in a splitting of each of the eigenvalues of a completely separated well into two neighboring ones, the splitting being proportional to $\exp{-L_{II}/L_c}$, while the corresponding  eigenfunctions are symmetric and antisymmetric combinations of the corresponding eigenfunctions of the separated left and right wells.

From the translation symmetry of the time independent equation Eq.~(\ref{MEDEQ}) it follows  that the eigenfunctions of the Hermitian operator $\hat{H}$ include the function    $\Psi_1= T_d/ dx $ corresponding to the eigenvalue $\lambda_1=0$ (it is easy to check inserting this function in the Eq.~(\ref{eigenfunctionequation})).
As it follows from Eq.~(\ref{trapezoid}) the eigenfunction $\Psi_1(x)$ is
\begin{eqnarray}\label{PsiAntisymmetric}
\Psi_1(x)=\frac{d T_d}{d x}= \frac{d \vartheta(x+x_+)}{d x} +\frac{d \vartheta(x_+-x)}{d x}
\end{eqnarray}
where two functions in the right-hand side are eigenfunctions of the left and right wells when the tunneling is ignored. This is an antisymmetric eigenfunction ($\Psi(-x) =-\Psi(x)$) corresponding to the eigenvalue $\lambda_1=0$ and hence the nearest eigenvalue $\lambda_0$ is negative and the corresponding eigenfunction is symmetric:
\begin{eqnarray}\label{PsiSymmetric}
\Psi_0(x)=\frac{d \vartheta(x+x_+)}{d x} - \frac{d \vartheta(x_+-x)}{d x}
\end{eqnarray}
while
\begin{eqnarray}\label{lambda0}
\lambda_0&=&-\kappa \left(\frac{d \vartheta(x+x_+)}{d x}\frac{d^2 \vartheta(x+x_+)}{d x^2}\right)\Big|_{x=0} \nonumber \\
&\propto& \exp{(-L/L_c)}
\end{eqnarray}

To express the eigenvalue $\lambda_0$ in terms of the differential resistivity we use the following reasoning.

According to Eq.~(\ref{current-bias}) the voltage-current characteristic of the stack with a MTED in it is
 \begin{eqnarray}
{\cal E}_d(j)=j\left<\rho[\Theta\big(T_d(j)\big)] \right>
\label{current-bias1}
\end{eqnarray}
 Differentiating the both sides of Eq.~(\ref{current-bias1}) and Eq.~(\ref{MEDEQ}) with respect to $j$  one finds  the differential resistivity of the magnetic stack with a MTED as
\begin{eqnarray}\label{DiffResisivityMTED}
\frac{d {\cal E}_d}{dj}=\langle\rho(T_d)\rangle+\langle\rho^{
'}_T(T_d)\frac{\partial T_d}{\partial j}\rangle.
\end{eqnarray}
and an equation for $\partial T_d/\partial j$
\begin{eqnarray}\label{TderivativeJeq}
\hat{H}\frac{\partial T_d}{\partial j} = 2j\langle\rho(T_d)\rangle
\end{eqnarray}
where the operator  $\hat{H}$ is given by Eq.~(\ref{Hamiltonian2}).

Expanding $\partial T_d/\partial j$ in the form of the eigenfunctions $\Psi_\nu$ (see Eq.~(\ref{eigenfunctionequation})) one finds
\begin{eqnarray}\label{dT/dj}
\frac{\partial T_d}{\partial j} =2j \sum_{\nu}\frac{\langle\rho(T_d)\Psi_\nu\rangle}{\lambda_\nu }\Psi_\nu
\end{eqnarray}

Inserting it in Eq.~(\ref{DiffResisivityMTED}) one finds
\begin{eqnarray}\label{MTEDCVCinEigenfunctions}
\frac{d {\cal E}_d}{dj}&=& \langle\rho(T_d)\rangle\nonumber \\
&+&2j_d^2\sum_\nu\frac{\langle\rho_T^{'}(T_d)\Psi_\nu\langle\rangle\rho(T_d)\Psi_\nu\rangle}{\lambda_\nu}\,.
\end{eqnarray}
Remembering that $\lambda_0$ is exponentially small and hence it gives the main contribution to the sum with respect to $\nu$ one finds
\begin{eqnarray}\label{lambda01}
\lambda_0=\frac{16 r (j_d^2 \rho_+)\int_{T_{min}}^{T_{max}}\rho[\Theta(T)]dT }{L^2<(dT_d/dx)^2> d {\cal E}_d/dj}
\end{eqnarray}


\begin{thebibliography}{99}
\bibitem{Giovanni}
The orientational phase transition in such a system induced by an external
magnetic field was considered in Ref.~\onlinecite{Asti}.
\bibitem{domain}
Electric domains in normal metals (see, e.g., the review Ref.~\onlinecite{Mints}) can also appear due to 1) structural  \cite{Barelko} and magnetic \cite{Landauer,Ross}  transitions, 2) a sharp dependence of the resistance on $T$ at low temperatures and magnetic breakdown \cite{Slutskin,TED,Chiang,Boiko,Abramov1}, 3) evaporation \cite{Atrazhev} and melting \cite{Abramov2}.
\bibitem{torquecurrent} Typical densities of critical currents needed for  the torque effect in point-contact devices are $10^8\div 10^9 \rm A/cm^2$  for the current  perpendicular to the layers (CPP). The record low  torque  critical current  in a nanopilar device was reported in Ref.~\onlinecite{Diao} (see Ref. ~\onlinecite{Ralph}). In our paper we consider the case in which the torque effect is absent that is the torque critical current is higher than $10^7 \rm A/cm^2$.
\bibitem{footnote2}
For the realistic experimental parameters $c_V\sim$~1~J/Kcm$^3$, $T_c^{(1)}\sim$~10$^2$~K, $\rho \sim$~10$^{-5}$~$\Omega$cm  and $j \sim$~$10^6\div 10^7$~A/cm$^2$ one has $\tau_T \sim 10^{-7}\div 10^{-5}$~s.
\end{thebibliography}
\end{document}